\documentclass[sigconf,natbib=false]{acmart}
\AtBeginDocument{%
  }

\setcopyright{acmlicensed}
\copyrightyear{2018}
\acmYear{2018}
\acmDOI{XXXXXXX.XXXXXXX}
\acmConference[ICCPS 2026]{ACM/IEEE International Conference on Cyber-Physical Systems (ICCPS)}{May 11-14, 2026}{Saint Malo, France}
\acmISBN{978-1-4503-XXXX-X/2018/06}

\usepackage{amsthm}
\usepackage{amsmath}
\usepackage{bm}
\usepackage{caption}
\usepackage{subcaption}
\usepackage{xcolor}
\usepackage[
  datamodel=acmdatamodel,
  style=acmnumeric,
  ]{biblatex}
  
\usepackage{xspace}

\usepackage{xcolor}

\theoremstyle{acmdefinition}
\newtheorem{remark}{Remark}
\newtheorem{assumption}{Assumption}

\newcommand{\bbN}{\mathbb{N}}  

\newcommand{\bbR}{\mathbb{R}}  

\newcommand{\bbRplus}{\ensuremath{\bbR^+}}

\newcommand{\calB}{\mathcal{B}}
\newcommand{\calC}{\mathcal{C}}
\newcommand{\calD}{\mathcal{D}}

\newcommand{\calT}{\mathcal{T}}

\newcommand{\eg}{e.g.,\xspace}
\newcommand{\ie}{i.e.,\xspace}

\newcommand{\restrict}[2]{\left.#1\right|_{#2}}

\def\fbsymbol{\circlearrowright\!}

\newcommand{\Aspace}{\mathcal{A}}
\newcommand{\Gspace}{\mathcal{G}}
\newcommand{\Uspace}{\mathcal{U}}
\newcommand{\Xspace}{\mathcal{X}}
\newcommand{\Yspace}{\mathcal{Y}}

\newcommand{\Uvec}{\bm{U}}
\newcommand{\Xvec}{\bm{X}}
\newcommand{\Yvec}{\bm{Y}}

\begin{document}

\title{$\partial$CBDs: Differentiable Causal Block Diagrams}


\author{Thomas Beckers}
\authornote{Authors are listed in alphabetical order; all authors contributed equally to this work.}
\affiliation{%
  \institution{Vanderbilt University}
  \city{Nashville}
  \country{USA}}
\email{thomas.beckers@vanderbilt.edu}

\author{J\'an Drgo\v{n}a}
\authornotemark[1]
\affiliation{%
  \institution{Johns Hopkins University}
  \city{Maryland}
  \country{USA}
}
\email{jdrgona1@jh.edu}

\author{Truong X. Nghiem}
\authornotemark[1]
\affiliation{%
  \institution{University of Central Florida}
  \city{Orlando}
  \country{USA}
}
\email{truong.nghiem@ucf.edu}


\begin{abstract}
Modern cyber-physical systems (CPS) integrate physics, computation, and learning, demanding modeling frameworks that are simultaneously composable, learnable, and verifiable. Yet existing approaches treat these goals in isolation: causal block diagrams (CBDs) support modular system interconnections but lack differentiability for learning; differentiable programming (DP) enables end-to-end gradient-based optimization but provides limited correctness guarantees; while contract-based verification frameworks remain largely disconnected from data-driven model refinement.
To address these limitations, we introduce differentiable causal block diagrams ($\partial$CBDs), a unifying formalism that integrates these three perspectives. Our approach (i) retains the compositional structure and execution semantics of CBDs, (ii) incorporates assume--guarantee (A--G) contracts for modular correctness reasoning, and (iii) introduces residual-based contracts as differentiable, trajectory-level certificates compatible with automatic differentiation (AD), enabling gradient-based optimization and learning. Together, these elements enable a scalable, verifiable, and trainable modeling pipeline that preserves causality and modularity while supporting data-, physics-, and constraint-informed optimization for CPS.

\end{abstract}



\maketitle

\section{Introduction}
Cyber-physical systems increasingly blend model-based design with data-driven components to achieve performance across diverse operating conditions~\cite{schmidt2018smart,friederich2021towards,ahmed2021machine,beckers2021online}. As these systems scale in complexity and autonomy, engineers face a three-way tension: compositionality to manage structural complexity, learnability to adapt models and controllers from data, and verifiability to ensure safety and correctness~\cite{varshney2017safety}. Today’s methodologies typically address one or two of these requirements, but rarely unify all three in a single, tractable workflow. Causal block diagrams (CBDs) remain the lingua franca of CPS design, see~\cite{gomesCausalblockDiagramsFamily2020}, due to their clear execution semantics and support for modular composition, yet they are not natively differentiable and thus only weakly connected to modern optimization and learning pipelines. Differentiable programming (DP) frameworks provide end-to-end gradients and sensitivity analysis, but operate over computational graphs whose semantics are geared toward numerical execution rather than correctness and compositional reasoning~\cite{innes2019differentiable}. Modern DP systems typically rely on automatic differentiation (AD) as their computational backbone, enabling efficient gradient propagation but offering limited structure for specifying or verifying formal contracts. Contract-based verification, especially via assume-guarantee specifications, provides scalable certification~\cite{benveniste2018contracts}, but is typically decoupled from learning, making it hard to enforce safety and performance constraints during data-driven optimization. This disconnect forces practitioners to choose between structured, certifiable models that are difficult to optimize and learning tools that lack robust assurances.

This paper introduces differentiable Causal Block Diagrams ($\partial$CBDs), a modeling and optimization framework that integrates CBD compositionality, assume-guarantee contract semantics, and differentiable programming into a single pipeline. In $\partial$CBDs, each block is equipped with explicit interfaces and time semantics, an execution map that may be continuous, discrete, or hybrid, and a set of differentiable contract residuals that formalize assumptions and guarantees. These residuals can quantify property satisfaction—such as passivity, gain bounds, and Lyapunov certificates—and can be composed across interconnections while preserving execution semantics. Crucially, $\partial$CBD compositions yield automatic-differentiation-executable graphs over finite horizons by rate lifting and loop unrolling, enabling gradients to propagate through physics-based models (e.g., ODE/DAE solvers, fixed-point and optimization layers) and data-driven components (e.g., neural networks) in a principled manner. Contracts thus become differentiable certificates that can be enforced or regularized during training, which allows learning to be guided by physics, safety, and performance constraints rather than solely by empirical loss.

\noindent\textbf{Related works.}
Automatic differentiation (AD) has become the computational backbone of modern machine learning and scientific computing, enabling efficient, machine-precision gradients through arbitrarily complex programs.
In machine learning, reverse-mode AD underpins deep learning frameworks such as TensorFlow, PyTorch, and JAX, supporting large-scale optimization of neural architectures~\cite{Abadi2016TensorFlow,Paszke2019pytorch,Bradbury2018JAX}.
In scientific computing and optimal control, AD extends to differential equations, providing sensitivities through numerical solvers~\cite{Andersson2019,SUNDIALS2005,hu2019difftaichi,chen2018neuralode}.
These advances have established AD as the key enabler of \emph{differentiable programming}, a paradigm that unifies simulation, learning, and optimization within a single computational graph~\cite{innes2019differentiable,baydin2018automatic,vanMerrienboer2018,Bolte2020}.
Building on this foundation, the proposed $\partial$CBD framework embeds AD semantics directly into the compositional structure of 
CBDs, enabling scalable, verifiable, and physically grounded learning in cyber-physical systems.

\noindent\textbf{Contributions.} First, we propose a differentiable extension of causal block diagrams that preserves compositional structure and execution semantics while enabling end-to-end gradient-based optimization. Second, we embed assume-guarantee contracts as differentiable residuals within the modeling pipeline, turning modular specifications into optimization-aware certificates that guide and constrain learning. Third, we develop a reduction from $\partial$CBDs to AD-executable graphs with well-defined sensitivity propagation across heterogeneous components, providing a foundation for scalable training and verification of CPS. Collectively, these elements establish a practical and principled basis for designing systems that are simultaneously composable, learnable, and verifiable.

\section{Methodology}

We introduce \emph{differentiable Causal Block Diagrams} ($\partial$CBDs), a formalism that unifies (i) the compositional modeling of CBDs, (ii) assume-guarantee (A--G) contracts for modular correctness, and (iii) differentiable programming (DP) for end-to-end learning.
By combining the structural clarity of CBDs, the compositional verification power of A--G contracts, and the scalability of automatic differentiation (AD) frameworks, our approach enables \emph{scalable, verifiable, and composable} modeling, learning, and optimization of complex cyber-physical systems.

In the proposed $\partial$CBD framework, each block is endowed with a differentiable contract encoding its physical, safety, or operational properties. CBD compositions (serial, parallel, or feedback) preserve these properties through contract composition and are translated to AD-executable directed acyclic graphs (DAGs). The resulting system behaves both as a causal diagram (for simulation and modular reasoning) and as a differentiable program (for learning and optimization). A--G contracts are embedded as smooth penalty or certificate functions within the computational graph, enabling formal correctness reasoning to coexist with gradient-based training.
Formally, the $\partial$CBD framework consists of five layers of abstraction:
\begin{enumerate}
    \item \textbf{Structural layer:} defines the block with contract as a fundamental unit in the causal block diagrams (Sec.~\ref{sec:CBD});
    \item \textbf{Semantic layer:} defines the composition rules and causal execution semantics for blocks (Sec.~\ref{sec:execution});
    \item \textbf{Contract layer:} assigns assume-guarantee specifications to blocks, ensuring compositional correctness (Sec.~\ref{sec:verification});
    \item \textbf{Differentiable layer:} exposes derivatives of the causal block diagrams via automatic differentiation (Sec.~\ref{sec:diff-cbd}).
    \item \textbf{Optimization layer:} enables verifiable gradient-based learning and optimization for causal block diagrams (Sec.~\ref{sec:optim}).
\end{enumerate}

Together, these layers yield a unified modeling and learning pipeline that preserves causality and compositional structure, exposes local gradients through AD-compatible primitives, and supports system-level verification via differentiable contracts at scale. This integration forms the methodological core of $\partial$CBDs.

\subsection{Causal Block Diagrams with Contracts} 
\label{sec:CBD}

The causal block diagram (CBD) is a widely adopted modeling formalism for practical model-based design of CPS \cite{gomesCausalblockDiagramsFamily2020}.
Perhaps the most prominent CBD implementation is Simulink \cite{denckla2005formalizing}, with recent extensions for contract-based semantics and refinement \cite{sunContractbasedSemanticsRefinement2023,zhangProvingSimulinkBlock2022}.
Similarly, Ptolemy II adopts the CBD as its core modeling paradigm, providing a flexible framework for composing heterogeneous models of computation within a unified block-based architecture \cite{Ptolemy2014}.
Our framework adopts the core CBD formalism commonly used in these implementations and the literature, described below.
The assume-guarantee (A--G) contract extension to CBD will be presented in Section~\ref{sec:verification}.

\subsubsection*{\bfseries Notations.}
An indexed lowercase letter, like $v_i$, denotes a variable, an uppercase letter, like $V$, is a set of variables $v_i$, a bold uppercase letter, like $\bm{V}$, is the vector concatenation of variables in $V$, and a calligraphic letter, like $\mathcal{V}$, is the value space of $\bm{V}$.

\subsubsection*{\bfseries Signals.}
We first define \emph{signals}, which are functions from time to values, that represent time-varying variables in a system.
The time domain is represented by non-negative real numbers, denoted by $\bbRplus$.
A signal can either be \emph{continuous-time} or \emph{discrete-time}.
\begin{definition}[Signal]
    A continuous-time signal \(s\) is a function from $\bbRplus$ to $S$, where \(S\) is the non-empty set of values of the signal.
    A discrete-time signal $s$ is a piece-wise constant continuous-time signal whose value only changes at discrete time instants $k \tau$, where the natural number $k \in \bbN$ is the time step and the positive real constant $\tau$ is the period, \ie $\forall t \in [k \tau, (k + 1) \tau), s(t) = s(k \tau)$.
\end{definition}
In the above definition, the value set $S$ can be any non-empty set, including finite sets of discrete values such as the Boolean set.
In practice, however, $S$ is often the real number set $\bbR$.
In this work, for simplicity, we assume that $S \equiv \bbR$.

\subsubsection*{\bfseries Blocks.}
\begin{figure}
    \centering
    \begin{subfigure}[b]{0.66\columnwidth}
         \centering
         \includegraphics[width=\textwidth]{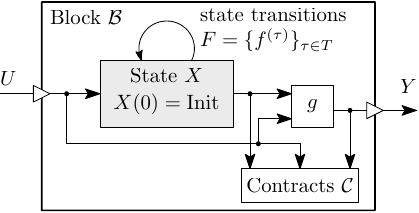}
         \caption{The components and semantics of a \emph{block}.}
         \label{fig:block-def}
    \end{subfigure}
    \hfill
    \begin{subfigure}[b]{0.30\columnwidth}
        \centering
        \includegraphics[width=\textwidth]{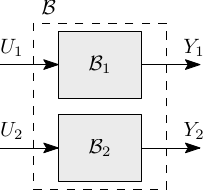}
        \caption{Parallel composition.}
        \label{fig:parallel-composition}        
    \end{subfigure}
    \hfill
    \begin{subfigure}[b]{0.56\columnwidth}
         \centering
         \includegraphics[width=\textwidth]{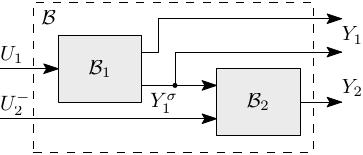}
         \caption{Serial composition.}
         \label{fig:serial-composition}
    \end{subfigure}
    \hfill
    \begin{subfigure}[b]{0.4\columnwidth}
        \centering
        \includegraphics[width=\textwidth]{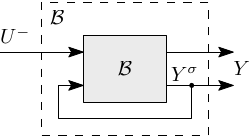}
        \caption{Feedback composition.}
        \label{fig:feedback-composition}        
    \end{subfigure}
    \caption{Illustrations of blocks and their compositions.}
    \Description{Illustrations of a block, the parallel composition, the serial composition, and the feedback composition.}
    \label{fig:block-compositions}
\end{figure}

In a CBD, a \emph{block} represents a component that typically transforms some input signals to some output signals with well-defined behaviors.
In our framework, each block can be associated with an \emph{assume-guarantee contract}, which specifies the behaviors of the block under given conditions.
Contracts are a key element in our framework for verification and for learning, specifying system properties such as safety requirements and physical priors that can be integrated into learning.
A block can be continuous-time, discrete-time, or hybrid (where both continuous-time and discrete-time behaviors exist), and can have multiple associated sampling periods.
We adopt the following formalization of a \emph{block with contract}, illustrated in Fig.~\ref{fig:block-def}.
\begin{definition}[Block with Contract]
\label{def:block}
A block is a tuple
\[
\calB = (U, X, Y, T, F, g, \mathrm{Init}, \calC),
\]
\begin{itemize}
\item $U = \{u_1,\dots,u_n\}$ is the set of \emph{input variables}, each denoting a signal that is consumed by the block.
\(\Uvec\) is the vector concatenation of all input variables and $\Uspace$ is its value space.

\item $X = \{x_1,\dots,x_p\}$ is the set of \emph{internal state variables}, each denoting a signal that is internal to the block.
$\Xvec$ is the vector concatenation of all state variables and $\Xspace$ is its value space.

\item $Y = \{y_1,\dots,y_m\}$ is the set of \emph{output variables}, each denoting a signal that is generated by the block.
$\Yvec$ is the vector concatenation of all output variables and $\Yspace$ is its value space.

\item $T$ is the \emph{time semantics} of the block, which is a set of unique sampling periods $T = \{\tau_1, \dots, \tau_q\}$, $\tau_i \in \bbRplus$, with the convention that a zero sampling period means continuous-time.  $T$ must contain at least one element.

\item $F = \{f^{(\tau)}\}_{\tau \in T}$ is the set of 
\emph{state transition functions} $f^{(\tau)}$ corresponding to the sampling periods $\tau$ in $T$.  Each $f^{(\tau)}$ only updates a non-empty subset $X^{(\tau)} \subseteq X$ of the state variables, and the set of $X^{(\tau)}$ is a partition of $X$, \ie $\cup_{\tau \in T} X^{(\tau)} = X$ and for any $\tau, \tau' \in T$, $X^{(\tau)} \cap X^{(\tau')} = \emptyset$.  The semantic of $f^{(\tau)}$ depends on its time semantic:
    \begin{itemize}
        \item If $\tau = 0$: all state signals in $X^{(0)}$ are continuous-time and have continuous values, and $f^{(0)}$ defines a differential equation: $\dot{\Xvec}^{(0)}(t) = f^{(0)} (\Xvec(t),\allowbreak \Uvec(t))$, that is well-defined with unique solution $\Xvec^{(0)}(t)$ for any initial condition.

        \item If $\tau > 0$: all state signals in $X^{(\tau)}$ are discrete-time and can be either continuous or discrete.  Their values are updated at each discrete time instant $k \tau$ by the difference equation: $\Xvec^{(\tau)}((k+1) \tau) = f^{(\tau)} (\Xvec(k \tau), \Uvec(k \tau))$.
    \end{itemize}

\item $g$ is the \emph{output function} that calculates the outputs $Y$ from the current states and inputs: $\Yvec(t) = g(\Xvec(t), \Uvec(t))$.  
It can represent an explicit equation from $\Xvec$ and $\Uvec$ to $\Yvec$ or an implicit equation of the form $\Yvec(t) = h(\Yvec(t), \Xvec(t), \Uvec(t))$, solved by a fixed-point solver.

\item $\mathrm{Init}$ is the initial condition of the states $X$, \ie $\Xvec(0) = \mathrm{Init}$.

\item $\calC$ is a set of \emph{contracts} satisfied by the block (see Sec.~\ref{sec:verification}).
\end{itemize}
\end{definition}

\begin{assumption}[Block]
    \label{asmp:block}
    A block 
    must satisfy the following: 
    \begin{itemize}
    \item It has at least one input or output, i.e., the sets $U$, $X$, and $Y$ can be empty but $U$ and $Y$ cannot be both empty.

    \item Its dynamics, specified by the state transition functions $f^{(\tau)}$ and the output functions $g_\tau$, are causal.

    \item Its dynamics are driven by time, not by events.
    \end{itemize}
\end{assumption}%

\begin{remark}
We make the following remarks. 
\begin{itemize}
    \item 
    If $U$ is empty then the block is a \emph{source block} that generates signals.
    If $Y$ is empty then it is a \emph{sink block} that only consumes signals.
    If $X$ is empty then the block is \emph{stateless} or \emph{static}; otherwise, the block is \emph{stateful} or \emph{dynamic}.

    \item A block may be hybrid, having both a continuous-time semantic ($\tau = 0$) and discrete-time semantics ($\tau > 0$).

    \item For the remainder of this paper, we will focus on continuous-valued states $X$. Event-driven dynamics with discrete state variables will be investigated in future research.  
\end{itemize}
\end{remark}

\subsection{Execution and Composition of CBDs}
\label{sec:execution}

Given a block $\calB$ and its input signals $\Uvec(t) = \{u_1(t),\allowbreak \dots,\allowbreak u_n(t)\}$, the block can be \emph{executed} to generate state signals $\Xvec(t) = \{x_1(t),\allowbreak \dots,\allowbreak x_p(t)\}$ and output signals $\Yvec(t) = \{y_1(t),\allowbreak \dots,\allowbreak y_m(t)\}$, for $0 \leq t \leq t_f$ where $t_f$ is the final time, by solving the \emph{causal} differential and difference equations resulting from $F$.  A brief description of the block execution algorithm is given below.
\begin{enumerate}
    \item Based on the discrete sampling periods in $T$, a sequence of time instants $0 = t_0, t_1, \dots, t_N = t_f$ is determined, where for each $t_i$ with $0 < i < N$, there exists $\tau \in T$ such that $t_i = k \tau$ for some $k \in \bbN$.

    \item Initial states and outputs are set: $\Xvec(0) = \mathrm{Init}$ and $\Yvec(0) = g (\Xvec(0),\allowbreak \Uvec(0))$.

    \item 
    For each interval $(t_i, t_{i+1}]$, the corresponding differential equation and difference equations are solved from the initial condition $\Xvec(t_i)$ to obtain the state values in the interval.
    Then, the output signals $Y$ are calculated by $g$.
\end{enumerate}
If the signals $\Xvec(t)$ and $\Yvec(t)$ are generated by the execution semantics of $\calB$ from the input signals $\Uvec(t)$, we write $(\Xvec(t), \Yvec(t)) \in \mathrm{Exec}\left(\calB, \Uvec(t)\right)$.
As a consequence, 
each block defines an \emph{input--output map} from the input signals $\Uvec(t)$ to the output signals $\Yvec(t)$, denoted by $\Yvec = \calB (\Uvec)$ using the same block notation.

\subsubsection*{\bfseries Compositions of blocks}
A CBD consists of blocks that are interconnected. 
A CBD can be ``flattened'' to an equivalent block by composing its blocks and their connections in a directed acyclic graph (DAG).
The remainder of this section formalizes a unilateral connection between two blocks and the fundamental composition rules of blocks.

\begin{definition}[Unilateral Connection]
    Given two blocks $\calB_i = (U_i, X_i, Y_i, T_i, F_i, g_i, \mathrm{Init}_i, \calC_i)$, $i = 1,2$, a \emph{unilateral connection} $\sigma_{1, 2}$ from $\calB_1$ to $\calB_2$ is a relation $\sigma_{1, 2} = \{(y_{1,i}, u_{2,j}) \,|\, y_{1,i} \in Y_1, u_{2,j} \in U_2\}$ satisfying:
    \begin{enumerate}
        \item The output and input in each pair $(y_{1,i}, u_{2,j})$ are compatible in terms of types and dimensions;
        \item Only one output can be connected to an input;
        \item $u_{2,j}(t) = y_{1,i}(t)$ for all $t \in \bbRplus$.
    \end{enumerate}
    The set of outputs and the set of inputs in the connection $\sigma_{1, 2}$ are denoted by $\sigma_{1, 2}^{\mathrm{out}} \subseteq Y_1$ and $\sigma_{1, 2}^{\mathrm{in}} \subseteq U_2$, respectively.
\end{definition}

A CBD is thus a composition of blocks and their connections.
\begin{definition}[Causal Block Diagram (CBD)]
\label{def:CBD}
    A CBD is a tuple $\calD = (\bm{\calB}, \Sigma)$ of blocks $\bm{\calB} = \{\calB_i\}_i$ and a set of unilateral connections between these blocks $\Sigma = \{\sigma_{i, j} \,|\, \calB_i \in \bm{\calB}, \calB_j \in \bm{\calB}\}$.
    The connections must satisfy that no more than one output can be connected to an input, but an output can be connected to more than one input.
\end{definition}

We now define the semantics of three elementary composition operators that can be used to compose a more complex CBDs.
The following notation will be required: if $f: X \mapsto Y$ is a function with a vector of output variables $Y$ and $Y' \subseteq Y$, then $\restrict{f}{Y'}$ denotes the function $f$ restricted to the subset $Y'$ of outputs, \ie $\restrict{f}{Y'} := \pi_{Y'} \circ f$ where $\pi_{Y'}$ is the projection operator from $Y$ to $Y'$.

A \emph{parallel composition} of two blocks, illustrated in Fig.~\ref{fig:parallel-composition}, simply combines the two blocks into a single block.
\begin{definition}[Parallel Composition]
\label{def:parallel-composition}
    Given two blocks $\calB_i = (U_i,\allowbreak X_i,\allowbreak Y_i,\allowbreak T_i,\allowbreak F_i,\allowbreak g_i,\allowbreak \mathrm{Init}_i,\allowbreak \calC_i)$, $i = 1,2$.
    A \emph{parallel composition} $\calB_1 \parallel \calB_2$ is a block $\calB = (U,\allowbreak X,\allowbreak Y,\allowbreak T,\allowbreak F,\allowbreak g,\allowbreak \mathrm{Init},\allowbreak \calC)$ with
    $U = U_1 \cup U_2$,
    $X = X_1 \cup X_2$,
    $Y = Y_1 \cup Y_2$,
    $T = T_1 \cup T_2$,
    $F = F_1 \cup F_2$,
    $g(\Xvec, \Uvec) = (g_1(\Xvec_1, \Uvec_1), g_2(\Xvec_2, \Uvec_2))$,
    $\mathrm{Init} = (\mathrm{Init}_1, \mathrm{Init}_2)$,
    and the composed contract $\calC$ to be presented in Section~\ref{sec:verification}.
\end{definition}

A \emph{serial composition} of two blocks connects some outputs of one block to some inputs of the other block, as illustrated in Fig.~\ref{fig:serial-composition}.
\begin{definition}[Serial Composition]
\label{def:serial-composition}
    Given two blocks $\calB_i = (U_i,\allowbreak X_i,\allowbreak Y_i,\allowbreak T_i,\allowbreak F_i,\allowbreak g_i,\allowbreak \mathrm{Init}_i,\allowbreak \calC_i)$, $i = 1,2$, and a connection $\sigma_{1, 2}$ from $\calB_1$ to $\calB_2$.
    Let $U_2^- = U_2 \setminus \sigma_{1, 2}^{\mathrm{in}}$ denote the inputs of $\calB_2$ not connected by the connection $\sigma_{1, 2}$, and let $Y_1^\sigma = \sigma_{1, 2}^{\mathrm{out}}$ denote the outputs of $\calB_1$ connected to $\calB_2$.
    A \emph{serial composition} $\calB_1 ; \calB_2$ is a block $\calB = (U,\allowbreak X,\allowbreak Y,\allowbreak T,\allowbreak F,\allowbreak g,\allowbreak \mathrm{Init},\allowbreak \calC)$ with
    \begin{itemize}
        \item $U = U_1 \cup U_2^-$, \, $X = X_1 \cup X_2$, \, $Y = Y_1 \cup Y_2$, \, $T = T_1 \cup T_2$,
        \item $F = F_1 \cup F_2^\sigma$, where $F_2^\sigma$ consists of the same state transition functions in $F_2$ restricted to the connected inputs determined by $\sigma_{1, 2}$.  Specifically,
        each $f_2^{(\tau)}(\Xvec_2, \Uvec_2)$ in $F_2$ becomes $f_2^{(\tau)}(\Xvec_2, (\Uvec_2^-, \restrict{g_1}{Y_1^\sigma}(\Xvec_1, \Uvec_1))) = f_2^{(\tau)}(\Xvec, \Uvec)$ in $F_2^\sigma$.
        \item
        $ g(\Xvec, \Uvec) = (g_1(\Xvec_1, \Uvec_1), g_2(\Xvec_2, (\Uvec_2^-, \restrict{g_1}{Y_1^\sigma}(\Xvec_1, \Uvec_1))))$
         reflects the connection from $\calB_1$ to $\calB_2$.
        It can also be represented as two steps: $\Yvec_1 = g_1(\Xvec_1, \Uvec_1)$ and $\Yvec_2 = g_2(\Xvec_2, (\Uvec_2^-, \pi_{Y_1^\sigma}(\Yvec_1)))$.
        \item $\mathrm{Init} = (\mathrm{Init}_1, \mathrm{Init}_2)$
        \item $\calC$ is the composed contract, to be presented in Section~\ref{sec:verification}.
    \end{itemize}
\end{definition}

An important and commonly used composition in CBDs is the \emph{feedback composition}, which feeds some outputs of a block $\calB$ back into some of its own inputs by a directed cycle, as illustrated in Fig.~\ref{fig:feedback-composition}.
A challenge of the feedback composition is the \emph{algebraic loop feedback}.
We first describe the concept of a \emph{direct feedthrough}.
If the output function $g$ of a block $\calB$ is such that an input $u_i$ is directly fed through $g$ to an output $y_j$, \ie the value of $u_i(t)$ directly affect the value of $y_j(t)$, then $\calB$ has a \emph{direct feedthrough} from $u_i$ to $y_j$. 
An \emph{algebraic loop} occurs when the connections between the outputs and inputs 
in a feedback composition form a closed cycle of direct feedthroughs,
\ie 
$u_{i}$ is directly fed through to $y_{j}$ and $y_{j}$ is connected to $u_{i}$.
In other words, an output of the block is driven by itself in the same time step.
Although an algebraic loop can be expressed as an algebraic equation, solved in the CBD execution, there are inherent difficulties in handling it in this way. 
An alternative technique for handling an algebraic loop is to break it by artificially inserting a stateful block, such as a unit-time delay or an integrator, into the loop.
While this might slightly alter the semantics of the original CBD, it is 
straightforward and is therefore often employed in practice, \eg in Simulink.
For simplicity, we will assume that all feedback compositions are algebraic loop-free.

\begin{assumption}
    There are no algebraic loops in the CBD.
\end{assumption}
Removing this assumption by explicitly handling algebraic loops will be investigated in future research.

\begin{definition}[Feedback Composition]
\label{def:feedback-composition}
    Given a block $\calB = (U,\allowbreak X,\allowbreak Y,\allowbreak T,\allowbreak F,\allowbreak g,\allowbreak \mathrm{Init},\allowbreak \calC)$ and a connection $\sigma_{\mathrm{fb}}$ from $\calB$ to itself.
    The composition is assumed to be algebraic loop-free.
    Let $U^- = U \setminus \sigma_{\mathrm{fb}}^{\mathrm{in}}$ denote the inputs of $\calB$ not connected by the connection $\sigma_{\mathrm{fb}}$ and let $Y^\sigma = \sigma_{\mathrm{fb}}^{\mathrm{out}}$ denote the outputs of $\calB$ connected by $\sigma_{\mathrm{fb}}$.
    The algebraic loop-free \emph{feedback composition} $\fbsymbol\calB$ is a block $\calB_\mathrm{fb} = (U_\mathrm{fb},\allowbreak X_\mathrm{fb},\allowbreak Y_\mathrm{fb},\allowbreak T_\mathrm{fb},\allowbreak F_\mathrm{fb},\allowbreak g_\mathrm{fb},\allowbreak \mathrm{Init}_\mathrm{fb},\allowbreak \calC_\mathrm{fb})$ with
    \begin{itemize}
        \item $U_\mathrm{fb} = U^-$, \, $X_\mathrm{fb} = X$, \, $Y_\mathrm{fb} = Y$, and $T_\mathrm{fb} = T$,
        \item $F_\mathrm{fb} = F^\sigma$ 
        consists of the same state transition functions in $F$ restricted to the connected inputs determined by $\sigma_{i, j}$.
        These functions are well-defined if no algebraic loops exist.
        \item$  g_\mathrm{fb}(\Xvec_\mathrm{fb},\allowbreak \Uvec_\mathrm{fb}) = g(\Xvec_\mathrm{fb},\allowbreak (\Uvec_\mathrm{fb},\allowbreak \restrict{g}{Y^\sigma}(\Xvec_\mathrm{fb},\allowbreak \Uvec_\mathrm{fb})))$
        is well-defined without algebraic loops.
        With algebraic loops, $g_\mathrm{fb}$ becomes an implicit equation, which is allowed in Definition~\ref{def:block} and can be solved by a fixed-point solver under mild conditions.
        \item $\mathrm{Init}_\mathrm{fb} = \mathrm{Init}$
        \item $\calC$ is the composed contract, to be presented in Section~\ref{sec:verification}.
    \end{itemize}
\end{definition}

The above compositions have several useful properties.
\begin{lemma}
Given blocks $\calB_1$, $\calB_2$, and $\calB_3$, and admissible connections between them, the following properties hold.
\begin{enumerate}
    \item $(\calB_1 ; \calB_2) ; \calB_3 = \calB_1 ; (\calB_2 ; \calB_3)$
    \item $\calB_1 \parallel \calB_2 = \calB_2 \parallel \calB_1$
    \item $(\calB_1 \parallel \calB_2) \parallel \calB_3 = \calB_1 \parallel (\calB_2 \parallel \calB_3)$.
\end{enumerate}
\end{lemma}

These properties follow directly from the associativity and commutativity of signal concatenation and substitution in the CBD execution semantics and are standard in 
CBD formalisms~\cite{gomesCausalblockDiagramsFamily2020,zhangProvingSimulinkBlock2022}.%
Using these fundamental compositions and their properties, any algebraic loop-free CBD $\calD = (\bm{\calB}, \Sigma)$ can be converted, or \emph{flattened}, into an equivalent block $\calB$ with the same execution semantics as the original CBD.
\begin{figure}
    \centering
    \includegraphics[width=\columnwidth]{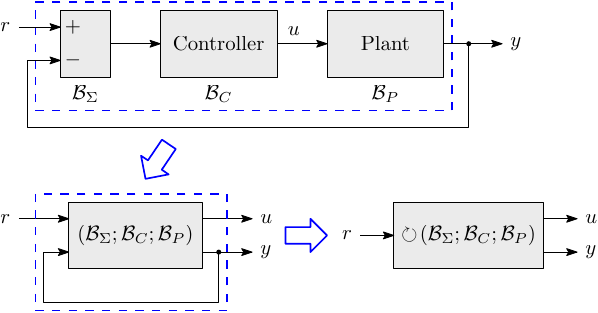}
    \caption{A standard feedback control loop (top) can be flattened into an equivalent block (bottom right) using the serial composition and feedback composition rules.}
    \Description{A standard feedback control loop can be flattened into an equivalent block using the serial composition and feedback composition rules.}
    \label{fig:composition-example}
\end{figure}
Fig.~\ref{fig:composition-example} demonstrates how a standard feedback control CBD is flattened into an equivalent block using the fundamental composition rules.
We first apply the serial composition rule on the feedforward path, which consists of the sum block $\calB_\Sigma$, the controller block $\calB_C$, and the plant block $\calB_P$, to obtain the composed block $(\calB_\Sigma ; \calB_C ; \calB_P)$.
We then apply the feedback composition rule on this composed block to flatten the original CBD into the block $\fbsymbol\!(\calB_\Sigma ; \calB_C ; \calB_P)$, which has the reference signal $r$ as input, and the control signal $u$ and plant output signal $y$ as outputs.

Furthermore, any subset of blocks of a CBD 
and their associated connections, which are also a CBD, can be replaced equivalently by their flattened block, called a \emph{subsystem}.
A subsystem can contain other subsystems. 
Consequently, this allows any CBD 
to be broken into nested subsystems.
This also allows a subsystem, represented by its flattened block, to be replaced by an approximate block, such as a machine learning based model.

\subsection{Contracts for CBDs}
\label{sec:verification}

CBDs support modular system construction, but reasoning about the correctness of interconnected blocks requires explicit specification of contractual guarantees. In this section, we introduce two complementary notions of contracts for CBDs. 
Assume--guarantee (A--G) contracts provide normative, system-wide specifications for compositional correctness reasoning, while residual-based contracts provide differentiable, quantitative certificates for realized executions and are used primarily for optimization and learning.

\subsubsection*{\bfseries Assume-Guarantee (A--G) Contracts for CBDs}

A--G contracts provide a formal framework for reasoning about system correctness through compositional specification and verification~\cite{benveniste2018contracts}, and have been successfully integrated into frameworks such as interface automata~\cite{de2001interface} and dynamical systems~\cite{saoud2021assume}. Building on these ideas, we extend the CBD formalism with A--G contract semantics, allowing for contract-based reasoning within block-structured models.

\begin{definition}[Assume--Guarantee Contract for a Block]
\label{def:ag-contract}
An assume--guarantee (A--G) contract for a block $\calB$ is a pair
\[
\calC^{\mathrm{AG}} = (\Aspace,\Gspace),
\]
where $\Aspace \subseteq \Uspace$ denotes the set of admissible input behaviors (assumptions), and $\Gspace \subseteq \Uspace \times \Yspace$ denotes the set of guaranteed input--output behaviors. The block $\calB$ is said to satisfy its contract if
\[
\forall U \in \Aspace : \quad (U, \calB(U)) \in \Gspace.
\]
\end{definition}

The composition of A--G contracts mirrors the structural composition of blocks and is formally defined by projecting and existentially quantifying internal signals while enforcing compatibility conditions between assumptions and guarantees.

For a set $\mathcal{S}\subseteq \Uspace\times\Yspace$, let $\pi_{U}(\mathcal{S})$ and $\pi_{Y}(\mathcal{S})$ denote its projections onto inputs and outputs, respectively. For serial connections $\sigma_{1,2}$, let $U_2^- = U_2\setminus \sigma_{1,2}^{\mathrm{in}}$ and $Y_1^\sigma=\sigma_{1,2}^{\mathrm{out}}$ as in Definition~\ref{def:serial-composition},
and let $\sigma_{1,2}(Y_1^\sigma)$ denote the mapping 
from signals $Y_1^\sigma$ to connected signals $\sigma_{1,2}^{\mathrm{in}}$.

\begin{definition}[A--G compatibility for a serial connection]
\label{def:ag-compat-serial}
Let $\calC_1^{\mathrm{AG}}=(\Aspace_1,\Gspace_1)$ and $\calC_2^{\mathrm{AG}}=(\Aspace_2,\Gspace_2)$ be A--G contracts for $\calB_1$ and $\calB_2$. 
The serial connection $\sigma_{1,2}$ is \emph{compatible} if every upstream behavior consistent with $\calC_1^{\mathrm{AG}}$ supplies a connected signal admissible to $\calB_2$, i.e.,
\[
\sigma_{1,2} \left(
\pi_{Y_1^\sigma}\!\left(\,\{(U_1,Y_1)\in \Gspace_1 \mid U_1\in \Aspace_1\}\right)
\right)
\;\subseteq\;
\pi_{\sigma_{1,2}^{\mathrm{in}}}(\Aspace_2).
\]
\end{definition}

\begin{definition}[A--G contract composition]
\label{def:ag-contract-composition}
Let $\calC_i^{\mathrm{AG}}=(\Aspace_i,\Gspace_i)$ be A--G contracts for blocks $\calB_i$.

\noindent\textit{Parallel:}
The parallel composition $\calB_1\parallel \calB_2$ has contract
\[
\calC_{\parallel}^{\mathrm{AG}} = (\Aspace_1\times \Aspace_2,\; \Gspace_1\times \Gspace_2),
\]
where $(\Aspace_1\times\Aspace_2)\!\subseteq\!(\Uspace_1\times \Uspace_2)$ and $(\Gspace_1\times\Gspace_2)\!\subseteq\! (\Uspace_1\times\Uspace_2)\times(\Yspace_1\times\Yspace_2)$.

\noindent\textit{Serial:}
Assume $\sigma_{1,2}$ is compatible in the sense of Definition~\ref{def:ag-compat-serial}. The serial composition $\calB_1;\calB_2$ has contract
\[
\calC_{;}^{\mathrm{AG}} = (\Aspace_{;},\Gspace_{;}),
\]
where $\Aspace_{;}\subseteq \Uspace_1\times \Uspace_2^-$ consists of external inputs for which the interconnection is admissible, and $\Gspace_{;}\subseteq (\Uspace_1\times \Uspace_2^-)\times(\Yspace_1\times \Yspace_2)$ is the set of external input/output behaviors obtained by existentially quantifying internal connection signals. Concretely, $(U_1,U_2^-,Y_1,Y_2)\in \Gspace_{;}$ if there exists a connected signal $v\in \pi_{\sigma_{1,2}^{\mathrm{in}}}(\Aspace_2)$ such that $(U_1,Y_1)\in \Gspace_1$ and $((U_2^-,v),Y_2)\in \Gspace_2$, with $v$ identified with $\pi_{Y_1^\sigma}(Y_1)$.

\noindent\textit{Feedback:}
Assume the feedback interconnection $\sigma_{\mathrm{fb}}$ is well-posed and compatible (analogous to serial compatibility). 
Then, feedback contracts are handled by unrolling feedback compositions into equivalent serial compositions, following the feedback execution semantics of CBDs, so that contract composition reduces to repeated application of the serial composition rule. This construction is equivalent to existentially quantifying internal feedback signals and projecting onto the remaining external variables.

\end{definition}

\begin{definition}[System-level A--G satisfaction]
A CBD $\calD=(\bm{\calB},\Sigma)$ satisfies its A--G contracts if each block $\calB_i$ satisfies $\calC_i^{\mathrm{AG}}=(\Aspace_i,\Gspace_i)$ under universal semantics and all interconnections in $\Sigma$ are compatible (i.e., guarantees imply downstream assumptions under the appropriate projections).
\end{definition}

\subsubsection*{\bfseries  Residual-based Contracts for CBDs}
To enable verifiable optimization and learning, we introduce residual-based contracts as quantitative certificates evaluated along realized executions.

\begin{definition}[Residual-Based Contract for a Block]
\label{def:residual-contract}
A residual-based contract for a block $\calB$ is defined by a mapping
\[
\calC^{\mathrm{res}} \colon 
\Xspace \times \Uspace \times \Yspace \times \mathbb{R}^{n_\theta}
\to \mathbb{R}^m,
\]
where $\theta \in \mathbb{R}^{n_\theta}$ parameterizes the contract. A realized execution $(X,U,Y)$ is said to satisfy the residual-based contract if
\[
\calC^{\mathrm{res}}(X,U,Y;\theta) \le 0
\]
holds element-wise. Residual satisfaction constitutes a \emph{sufficient certificate} that the realized input--output behavior $(U,Y)$ satisfies the corresponding guarantee for the given execution.
\end{definition}

Residual-based contracts are composed at the structural level according to the CBD execution semantics, by aggregating local residuals and substituting connected outputs into inputs.

\begin{definition}[Residual contract composition]
\label{def:res-contract-composition}
Consider residual-based contracts $\calC^{\mathrm{res}}_i(X_i,U_i,Y_i;\theta_i)\le 0$ for blocks $\calB_i$.

\noindent\textit{Parallel:}
The parallel composition $\calB_1\parallel \calB_2$ has residual contract given by concatenation
\[
\calC^{\mathrm{res}}_{\parallel}(X,U,Y;\theta)
=
\begin{bmatrix}
\calC^{\mathrm{res}}_1(X_1,U_1,Y_1;\theta_1)\\
\calC^{\mathrm{res}}_2(X_2,U_2,Y_2;\theta_2)
\end{bmatrix}
\le 0,
\]
with $X=(X_1,X_2)$, $U=(U_1,U_2)$, $Y=(Y_1,Y_2)$, and $\theta=(\theta_1,\theta_2)$.

\noindent\textit{Serial:}
For a serial interconnection $\sigma_{1,2}$, define $U_2^- = U_2\setminus \sigma_{1,2}^{\mathrm{in}}$ and $Y_1^\sigma=\sigma_{1,2}^{\mathrm{out}}$. The serial composition $\calB_1;\calB_2$ has residual contract
\[
\calC^{\mathrm{res}}_{;}(X,U,Y;\theta)
=
\begin{bmatrix}
\calC^{\mathrm{res}}_1(X_1,U_1,Y_1;\theta_1)\\
\calC^{\mathrm{res}}_2\!\big(X_2,\,(U_2^-,\,Y_1^\sigma),\,Y_2;\theta_2\big)
\end{bmatrix}
\le 0\text.
\]

\noindent\textit{Feedback:}
For a feedback interconnection $\sigma_{\mathrm{fb}}$, let $U^- = U\setminus \sigma_{\mathrm{fb}}^{\mathrm{in}}$ and $Y^\sigma=\sigma_{\mathrm{fb}}^{\mathrm{out}}$. The feedback composition $\fbsymbol\calB$ has residual contract
\[
\calC^{\mathrm{res}}_{\fbsymbol}(X,U^-,Y;\theta)
=
\calC^{\mathrm{res}}\!\big(X,\,(U^-,\,Y^\sigma),\,Y;\theta\big)
\le 0,
\]
where $Y^\sigma$ denotes the feedback outputs substituted into the corresponding feedback inputs.
\end{definition}

\begin{definition}[System-level residual satisfaction]
A CBD $\calD = (\bm{\calB}, \Sigma)$ satisfies its residual-based contracts for a realized execution if, for each block $\calB_i$, one has $\calC^{\mathrm{res}}_i(X_i,U_i,Y_i;\theta_i)\le 0$, and all interconnections respect admissible input domains.
\end{definition}




\begin{remark}[Residual contracts as exact A--G contracts]
\label{rem:res-AG-equivalence}
In special cases, a residual-based contract admits an exact interpretation as an A--G contract when it encodes a necessary and sufficient condition for the guarantee set $\Gspace$ and is enforced by construction for all admissible executions. In this setting, the residual depends only on parameters (or structural properties) of CBD and not on specific trajectories, yielding a universally quantified guarantee. Otherwise, residual-based contracts are trajectory-dependent and provide sufficient certificates only for the realized execution. 
We demonstrate representative structural property cases in Examples 1 and 3 in Section~\ref{sec:examples}.
Characterizing and exploiting this equivalence more broadly is a direction for future work.
\end{remark}

\subsection{Differentiable Causal Block Diagrams}
\label{sec:diff-cbd}

Automatic differentiation (AD) relies on Jacobian--vector products (JVPs) and vector--Jacobian products (VJPs). For a differentiable map $F:\mathbb{R}^n\to\mathbb{R}^m$ with Jacobian $J_F= \frac{\partial F}{\partial x}$, these are defined as
\begin{equation}
\label{eq:JVP_VJP}
\mathrm{JVP}(v)=J_Fv,\qquad \mathrm{VJP}(\bar v)=J_F^\top\bar v.
\end{equation}
Here, $v$ denotes a tangent (perturbation) direction in the input space, and $\bar{v}$ an adjoint (co-vector) in the output space. The JVP computes directional derivatives in the forward mode (sensitivity propagation), while the VJP computes adjoint sensitivities in the reverse mode (gradient backpropagation).
We assume throughout that all maps are defined on open sets of appropriate Euclidean spaces; statements ``a.e.'' refer to almost-everywhere differentiability.

\subsubsection*{\bfseries Differentiable Blocks.}  We now define the conditions required for a block with contract (Definition~\ref{def:block}) to be differentiable.

\begin{definition}[Differentiable Block]
\label{def:diff_block}
Consider a block
$
\calB=(U,\allowbreak X,\allowbreak Y,\allowbreak T,\allowbreak F,\allowbreak g,\allowbreak \mathrm{Init},\allowbreak \calC),
 $
where $\calC$ is a (possibly empty) set of contract annotations. 
We say that $\calB$ is \emph{(piecewise) differentiable} on $\Uspace\times\Xspace$ if Definition~\ref{def:block} holds and:
\begin{itemize}
\item \textit{Continuous-valued state variables.} $X$ are continuous-valued.

\item \textit{State transition functions $F$.} Each $f^{(\tau)}:\Xspace\times\Uspace\to\Xspace^{(\tau)}$ is locally Lipschitz and hence, by Rademacher’s theorem,  differentiable a.e., with bounded Jacobian on compact subsets.

\item \textit{Output map $g$.} The map $g:\Xspace\times\Uspace\to\Yspace$ is locally Lipschitz and hence differentiable a.e., with a declared policy at isolated non-smooth points (e.g., subgradient or smoothing).

\item \textit{Contracts $\calC$.} A--G contracts impose semantic constraints and do not participate in differentiation.
Residual-based contracts are given by locally Lipschitz residuals $\calC^{\mathrm{res}}_j(X,U,Y;\theta)$  that are differentiable a.e.\ and provide generalized gradients at non-differentiable points via an explicit AD policy (e.g., subgradients or smoothing).
\end{itemize}
\end{definition}

Under these conditions, each differentiable block defines an AD-compatible primitive that supports both forward- and reverse-mode sensitivity propagation.
Many engineering and machine learning blocks are piecewise $C^1$.
Depending on the types of state transition functions and output maps, several differentiable components can be considered. We enumerate the most important modalities below.

    \textit{i) Discrete-time dynamics.} 
    For a difference equation  $f^{(\tau)}$
   with $\tau > 0$, the Jacobian actions are computed via unrolling the discrete transition over the finite horizon. Here, the VJP is equivalent to the \emph{backpropagation through time} (BPTT) algorithm~\cite{werbos1990bptt,puskorius1994truncated}.

      \textit{ii) Continuous-time dynamics.}
    For a differential equation $f^{(\tau)}$ with $\tau = 0$, the Jacobian actions are computed via the chosen sensitivity method, either by unrolling the numerical integration and applying BPTT or via the continuous-time adjoint sensitivity method~\cite{dreyfus1962adjoint,CaoAdjointDAEs2003}.
    In case $f^{(\tau)}$ is a neural network, this is equivalent to the neural ordinary differential equations (NODEs) method~\cite{chen2018neuralode}.

      \textit{iii) Machine learning models.}
For static data-driven components, the block output $Y = g_\Theta(X,U)$ depends on learnable parameters~$\Theta$.
When the model is composed of differentiable primitives (e.g., linear, activation, and normalization layers), the mapping $g_\Theta$ is differentiable a.e.\ in $(X,U,\Theta)$, and JVP/VJP operations are provided natively by automatic differentiation frameworks~\cite{baydin2018automatic,Bolte2020,vanMerrienboer2018}.

      \textit{iv) Fixed-point solvers.}
If the output block internally solves an implicit equation of the form
\(
Y^\star = \Phi(Y^\star, U),
\)
assume $(I - \partial_Y \Phi)$ is nonsingular (e.g., contraction holds).
By the implicit function theorem, the local sensitivity of the equilibrium map $Y^\star(U)$ satisfies
$ 
(I - \partial_Y \Phi)\,\delta Y^\star = \partial_U \Phi\,\delta U,
$
whose solution defines the directional derivative
$
\delta Y^\star = J_{Y^\star,U}\,\delta U,$
with
$
J_{Y^\star,U} = (I - \partial_Y \Phi)^{-1}\partial_U \Phi.
$
This expression corresponds directly to the JVP of the equilibrium map, evaluated via the same linear solve used at runtime.
This allows a wide range of fixed-point or equilibrium solvers to be incorporated as differentiable blocks within the $\partial$CBD formalism~\cite{Sambharya2024,King2024,NEURIPS2019_01386bd6}.

\textit{v) Optimization layers.}
When the block output $Y^\star(U)$ is defined implicitly as the solution of a parametric optimization problem
\[
Y^\star(U)=\arg\min_Y \ \ell(Y,U)\quad\text{s.t.}\quad h(Y,U)=0,
\]
the solution is characterized by the KKT conditions
$
K(Y^\star,\lambda^\star,U)=0,
$
where $K=[\nabla_Y L(Y^\star,\lambda^\star,U);\;h(Y^\star,U)]$ and $L$ denotes the Lagrangian.
Under LICQ and strong regularity, the implicit function theorem yields the sensitivity relation
$
\partial_{(Y,\lambda)}K
\begin{bmatrix}\delta Y^\star\\ \delta\lambda^\star\end{bmatrix}
+
\partial_U K\,\delta U = 0,
$
which defines the Jacobian actions via the same linear solve used at runtime, enabling differentiable optimization layers within the $\partial$CBD formalism~\cite{Optnet2017,diffcp2019,tang2024,osti_10337537,jaxopt_implicit_diff,besancon2023diffopt,cvxpylayers2019}.

\textit{vi) Integer and logic ports.}
    For discrete or logical signals, a differentiability policy must be declared, such as a stop-gradient, straight-through estimator (STE), or smooth relaxation scheme, ensuring well-defined a.e.\ JVP/VJP behavior~\cite{bengio2013estimating,NEURIPS2022_43d8e5fc,L2OMINLP2025}.

\subsubsection*{\bfseries Differentiable Causal Block Diagrams ($\partial$CBD)} Now we have all the necessary components to define differentiable CBD.

\begin{definition}[Differentiable Causal Block Diagram]
\label{def:diff_CBD}
The CBD $\calD = (\bm{\calB}, \Sigma)$, cf.~Definition~\ref{def:CBD}, is \emph{differentiable} on $ U$ if:
\begin{enumerate}
\item Each block $\mathcal{B}_i \in \bm{\calB}$ is differentiable on its domain as per Definition~\ref{def:diff_block}.
\item The interconnection $\Sigma$ is type-consistent and feedback admissible.
\item The CBD input--output map $\calD: \Uspace \to \Yspace$, obtained by executing the CBD composition (Definition~\ref{def:CBD}), is differentiable a.e. and supports AD via local JVP/VJP composition.
\end{enumerate}
\end{definition}

\textit{Reduction to a differentiable directed acyclic graph (DAG):}
AD engines operate on DAGs. Any differentiable CBD can be reduced to such a DAG for a given execution horizon by:
\begin{enumerate}
\item \textit{Rate lifting:} lift all time bases $T_i$ to a common hyperperiod so evaluations are synchronous.
\item \textit{Parallel composition (branch expansion):} split $\mathcal{B}_1\parallel\mathcal{B}_2$ into independent subgraphs and concatenate ports.
\item \textit{Feedback composition:} unroll each feedback loop for a finite number of steps.
\item \textit{Event flattening (hybrid):} encode mode switches with declared differentiability policy (e.g., relaxed gates).
\item \textit{Topological sort:} sort the resulting acyclic graph to obtain an execution schedule $\mathcal{S}$ (evaluation trace) compatible with forward and reverse AD.
\end{enumerate}
This produces a rate-lifted, branch-expanded, feedback-unrolled DAG whose CBD composite map $\calD$ is differentiable and AD-ready.
Let $J_{\calD} = \partial F/\partial U$ denote the total Jacobian of $\calD$. AD never computes the full Jacobian $J_{\calD}$; instead, it applies local Jacobian actions along $\mathcal{S}$.
A practical pipeline is:
\begin{enumerate}
\item Build the differentiable CBD and perform rate lifting, branch expansion, loop unrolling, and event flattening;
\item Execute the sorted DAG to record the evaluation trace $\mathcal{S}$;
\item Run AD sweeps (JVP for sensitivity analysis, VJP for learning) along $\mathcal{S}$.
\end{enumerate}

This formalism, thereby, allows the integration of physics-based blocks (such as ODE integrators, optimization layers) and data-driven blocks (such as neural networks) in a unified, \emph{differentiable}, and \emph{certifiable} computational graph that is suitable for implementation on modern AD frameworks.

\subsection{Verifiable Optimization and Learning with Differentiable CBDs}
\label{sec:optim}

Building on the composability, differentiability, and verifiability properties of the proposed $\partial$CBD framework, we now formulate a general optimization problem that unifies learning, modeling, optimization, and control within CPS.  
This formulation leverages the differentiable execution semantics of $\partial$CBDs to enable gradient-based learning and optimization, while incorporating residual-based contracts as differentiable constraints that enforce correctness, safety, and performance requirements.

\subsubsection*{\bfseries General $\partial$CBD Optimization Problem.}
We consider parameterized diagrams $\calD(\theta) = (\bm{\calB}(\theta),\Sigma)$ over a finite horizon $\calT = [0,t_f]$ (continuous time) or $\calT = \{0,\dots,N\}$ (discrete time). Let $(X,U,Y)$ denote the trajectories generated by executing $\calD(\theta)$ per Section~\ref{sec:CBD}. The CBD constrained optimization problem is cast as:
\begin{subequations}
\label{eq:verifiable-opt}
\begin{align}
\min_{\theta}\quad & \mathcal{J}\left(\Yvec(t), \Uvec(t), \Xvec(t);\theta\right) \;+\; w_{\mathrm{reg}}\,\mathcal{R}(\theta) \\
\text{s.t.}\quad 
& (\Xvec(t), \Yvec(t)) \in \mathrm{Exec}\left(\calD(\theta), \Uvec(t)\right), \quad \text{where} \\
& \text{(dynamics)}\;\; 
\begin{cases}
\dot{X}^{(0)}(t) = f^{(0)}\!\left(\Xvec(t),\Uvec(t);\theta\right), & t\in\calT,\\
X^{(\tau)}_{k+1} = f^{(\tau)}\!\left(X_k,U_k;\theta\right), & \tau>0,\; k\in\calT,
\end{cases}\\
& \text{(algebraic outputs)}\;\; \Yvec(t) = g\left(\Xvec(t), \Uvec(t);\theta\right), \\
& \text{(contracts)}\;\; \calC^{\text{res}}\left(\Xvec(t), \Uvec(t), \Yvec(t);\theta\right) \;\le\; 0,\quad  \forall t\in\mathcal{T}. 
\end{align}
\end{subequations}
\noindent Here, $\mathcal{J}$ encodes task performance (e.g., tracking, cost, energy, robustness), $\mathcal{R}$ is a regularizer, and $\mathrm{Exec}(\cdot)$ denotes the forward execution semantics of the CBD under the serial/parallel/feedback compositions. 
Residual contracts from Section~\ref{sec:verification} instantiate differentiable constraints
\(
\calC^{\text{res}}(\theta)\le 0
\)
whose gradients integrate into training or certification objectives. Thus, performance and safety (e.g., passivity, Lyapunov stability, gain bounds) can be optimized jointly within the same AD pipeline. When composing blocks, contract residuals sum (or take maxima) across the DAG, preserving modular verification with gradient access.

\paragraph{Scope and generality.}
The optimization problem~\eqref{eq:verifiable-opt} defines a general and composable framework that encompasses a broad range of modeling, learning, optimization, and control tasks.  
Depending on the choice of blocks, contracts, and objective $\mathcal{J}$, it can represent:
\begin{itemize}
    \item \textit{Constrained optimization}, encompassing a broad family of optimization problems, such as convex, nonlinear, deterministic, stochastic, or bilevel formulations, expressed as compositions of CBD blocks.
    \item \textit{Constrained system identification}, where $\theta$ parameterizes unknown dynamics or surrogate models, and contracts encode physical priors or stability constraints;
    \item \textit{Optimal control and policy optimization}, where $\theta$ parameterizes a feedback policy or optimal sequence of control actions. 
    \item \textit{Control co-design}, in which system model and control policy parameters are optimized jointly within the same CBD.
    \item \textit{Constrained machine learning and scientific ML}, where the CBD encodes general or physics-informed machine learning models subject to safety or feasibility guarantees.
\end{itemize}
Hence, this formulation admits joint optimization of heterogeneous subsystems, providing a unified foundation for end-to-end, verifiable design of complex cyber-physical systems.

\paragraph{Differentiability of the CBD-constrained optimization problem.}
Under the differentiability conditions of Definition~\ref{def:diff_CBD}, all constraints in~\eqref{eq:verifiable-opt} expose local JVP/VJP operations, enabling gradient-based optimization. 
Each block and contract contributes local Jacobian actions, making the entire diagram AD-ready and supporting joint, end-to-end optimization across heterogeneous subsystems.
Gradients of the objective and contract residuals are obtained by reverse-mode AD on the rate-lifted and loop-unrolled DAG representation of the CBD (Section~\ref{sec:diff-cbd}).  
Local JVP/VJP operations compose seamlessly through serial, parallel, and feedback connections, enabling scalable, memory-efficient backpropagation and verification across multiple domains, time scales, and decision layers.

\subsubsection*{\bfseries Gradient-based Optimization and Learning of $\partial$CBDs.}
To solve the verifiable optimization problem~\eqref{eq:verifiable-opt}, we adopt a reformulation that enables end-to-end training and verification within the AD pipeline.  
Contract and inequality constraints are relaxed into smooth penalties or barriers, yielding the unconstrained objective
\begin{equation}
\label{eq:penalty-barrier}
\min_{\theta}  
 \mathcal{L}(\theta)
= \mathcal{J}(Y,U,X;\theta)
+ w_{\mathrm{reg}}\,\mathcal{R}(\theta)
+ w_{\mathcal{C}}\,\Phi_{\beta}\big(C(X,U,Y;\theta)\big), 
\end{equation}
where, $\Phi_{\beta}$ denotes a penalty or barrier function, and $w_{\mathcal{C}}>0$ balances task performance and contract satisfaction.  
Gradients of~\eqref{eq:penalty-barrier} are obtained by reverse-mode AD on the computational graph of $\partial$CBD, enabling scalable optimization via gradient-based solvers.

\begin{remark}[Strict feasibility]
Feasibility restoration layers, such as differentiable optimization layers and fixed-point solvers~\cite{cvxpylayers2019,diffcp2019,donti2021dc3,L2OMINLP2025}  can enforce contract satisfaction with strict guarantees.  
Thanks to the compositional structure of  $\partial$CBD, these mechanisms can be applied locally per block or globally at the system level.
\end{remark}

\begin{remark}[Alternative optimization backends]
While we use a penalty--barrier formulation~\eqref{eq:penalty-barrier} for AD-compatible training, the $\partial$CBD framework is agnostic to the optimization backend.  
Augmented Lagrangian, operator-splitting, or classical NLP solvers (e.g., IPOPT) can be integrated and exploit gradients supplied by AD, as in CasADi~\cite{Andersson2019} or JAX~\cite{jaxopt_implicit_diff}.
\end{remark}

\begin{remark}[Scalability]
Scalability in $\partial$CBDs depends on three largely orthogonal factors: execution and differentiation of the CBD computational graph, A--G verification under the chosen formal methods, and numerical optimization of~\eqref{eq:penalty-barrier}. All three benefit from the compositional structure of CBDs, enabling modular decomposition and backend-specific scalability. A systematic study of scalability, spanning theoretical complexity bounds and empirical behavior across different verification and optimization backends, is an important direction for future work.
\end{remark}

\section{Examples}
\label{sec:examples}

We present three case studies illustrating how $\partial$CBDs compile compositional CBD models with contracts into AD-ready graphs for gradient-based design\footnote{The anonymized code is available at \url{https://zenodo.org/records/18226276}.}. The examples cover (i) stability-by-construction gain tuning, (ii) trajectory-level Lyapunov certificate learning for a nonlinear system, and (iii) stable-by-construction Deep Koopman identification.
All examples follow the same structure: (i) specify a CBD by composing blocks, (ii) specify either an A--G or a residual contract, (iii) solve a contract-constrained optimization problem via AD through the compiled $\partial$CBD graph, and (iv) report task performance and contract satisfaction.
We implement all examples using \texttt{PyTorch}~\cite{Paszke2019pytorch} and \texttt{Neuromancer}~\cite{Neuromancer2023}.

\subsection{Example 1: Tuning of a Scalar Feedback Gain with Input to State Stability Certificate}
\label{sec:example-siso}

The purpose of this introductory example is to demonstrate the utilization of our $\partial$CBD framework in stability-certified tuning of a scalar feedback gain.
We consider here a simple closed-loop system using three blocks as illustrated in Figure~\ref{fig:composition-example}.  

\paragraph{$\partial$CBD representation} 
We assume the plant model with a bounded additive disturbance $|w_k|\le w_{\max}$ represented as a stateful block $\calB_P$ with input $U_P=\{u,w\}$, state $X_P=\{x\}$ and output $Y_P=\{y\}$, sampling period $T_P=\{\tau\}$, and state transition and output maps
\begin{subequations}
\label{eq:siso}
    \begin{align}
& x_{k+1} = f^{(\tau)}_P(x_k,u_k,w_k;a,b)
= a\,x_k + b\,u_k + w_k, \\
&  y_k = g_P(x_k,u_k,w_k)=x_k,
\end{align}
\end{subequations}
parametrized by $a\in\mathbb{R}$, $b>0$.
The static proportional controller
is represented as a stateless controller block $\calB_C$ with input $U_C=\{e\}$, output $Y_C=\{u\}$, and algebraic output map
$ 
u_k = g_C(e_k;\kappa) = \kappa\,e_k,
$
parameterized by the tunable gain $\kappa\in\mathbb{R}$.
The tracking error is represented as a stateless sum block $\calB_\Sigma$ with inputs $U_\Sigma=\{r,y\}$, output $Y_\Sigma=\{e\}$, and output map
$ 
e_k = g_\Sigma(r_k,y_k) = r_k - y_k.
$

Applying serial composition to $\calB_\Sigma$, $\calB_C$, and $\calB_P$, and then closing the loop by feeding $y$ back into $\calB_\Sigma$, yields the closed-loop block
\[
\calB_{\mathrm{cl}}
=\;
\fbsymbol\!\big(\calB_\Sigma \; ;\; \calB_C \; ;\; \calB_P\big),
\]
which is governed (for regulation $r_k\equiv 0$) by
$ 
y_{k+1} = (a - b \kappa)\,y_k + w_k.
$

\paragraph{Stability contract.}
In the scalar case, exponential stability of a linear system is equivalent to $|a - b \kappa| < 1$. A residual contract 
\[
\calC^{\text{res}}_{\mathrm{stab}}(\kappa,a,b)
\;=\;
\big|a - b \kappa\big| - (1 - \epsilon),
\qquad \epsilon \in (0,1),
\]
attached to $\calB_{\mathrm{cl}}$ encodes this property. The stability contract  with parameters $(\kappa,a,b)$ is satisfied if and only if
$ 
\calC^{\text{res}}_{\mathrm{stab}}(\kappa,a,b) \;\le\; 0,
$ 
i.e., $|a - b \kappa| \le 1 - \epsilon$, which enforces a strict stability margin for the closed-loop CBD. Under bounded disturbances $|w_k|\le w_{\max}$, this guarantees input-to-state stability (ISS) and convergence toward a disturbance-dependent neighborhood of the origin.

\paragraph{Contract-guided gain tuning.}
Let $\calT = \{0,\dots,N\}$ be a finite horizon. Executing the closed-loop block $\calB_{\mathrm{cl}}$ with parameter $\kappa$ over $\calT$ yields trajectories
$(X,U,Y) = \mathrm{Exec}(\calB_{\mathrm{cl}}(\kappa),W)$ for a disturbance realization $W=\{w_k\}_{k\in\calT}$,  where $Y=\{y_k\}_{k\in\calT}$ and $U=\{u_k\}_{k\in\calT}$.  
We tune the scalar gain $\kappa$ by solving the CBD-constrained optimization
\begin{subequations}
 \vspace{-0.3cm}
\label{eq:ex1-k-opt}
\begin{align}
\min_{\kappa\in\mathbb{R}} \quad
& \mathcal{J}(\kappa)
:= \sum_{k=0}^{N} \big( \|y_k\|^2 + \lambda_u \|u_k\|^2 \big), \\
\text{s.t.}\quad
& (U,Y) \in \mathrm{Exec}\big(\calB_{\mathrm{cl}}(\kappa), W\big), \\
& \calC^{\text{res}}_{\mathrm{stab}}(\kappa,a,b) \;\le\; 0.
\end{align}
\end{subequations}
This is a scalar instance of the general $\partial$CBD verifiable optimization problem~\eqref{eq:verifiable-opt} with parameter $\theta=\kappa$ and a single stability contract.  
In this example, instead of a penalty--barrier method~\eqref{eq:verifiable-opt}, we enforce the stability contract using a \emph{projected gradient step}, which clamps $\kappa$ to the admissible interval $\{\,|a-b\kappa|\le 1-\epsilon\,\}$ after each update.
Hence, this example demonstrates the special case described in Remark~\ref{rem:res-AG-equivalence}, where the residual
$\calC^{\text{res}}_{\mathrm{stab}}(\kappa,a,b)\le 0$ encodes a necessary and sufficient condition for exponential stability of the closed-loop scalar system, and is enforced by design through projection onto the admissible parameter set. 
Consequently, residual satisfaction is equivalent to satisfaction of the corresponding A--G stability contract under universal quantification.

\begin{figure}[ht]
    \centering
\includegraphics[width=0.99\columnwidth]{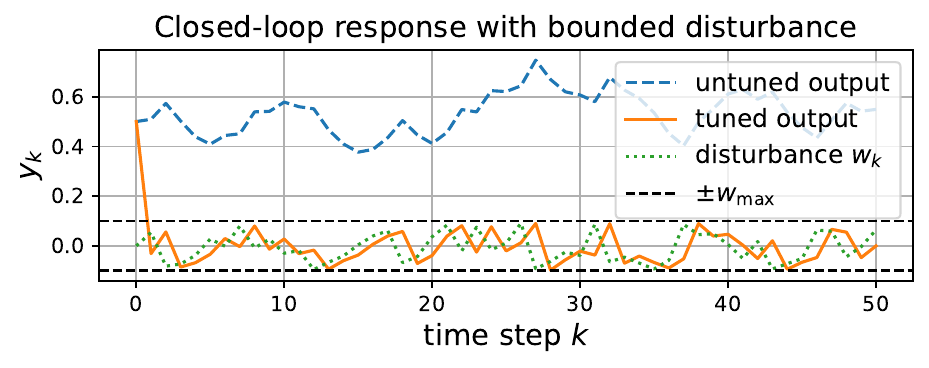}
        \vspace{-0.6cm}
    \caption{Closed-loop evolution of tuned vs untuned system.}
    \Description{A plot of closed-loop evolution of tuned versus untuned system.}
    \label{fig:siso_cl}
\end{figure}

\begin{figure}[ht]
    \centering
\includegraphics[width=0.99\columnwidth]{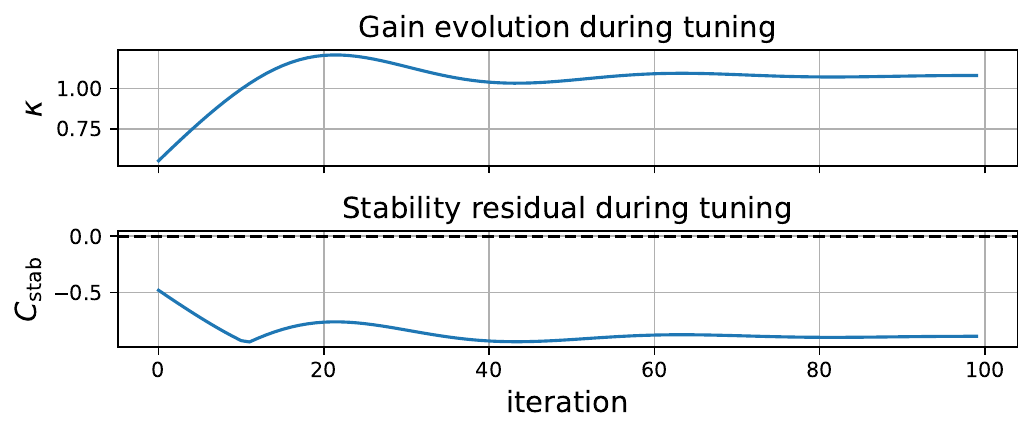}
    \vspace{-0.2cm}
    \caption{Training evolution of gain and contract residual.}
    \Description{Plots of training evolution of gain and contract residual.}
    \label{fig:siso_train}
    \vspace{-0.5cm}  
\end{figure}

\paragraph{Simulation results.}
To illustrate the behavior of the contract-guided tuning, we simulate the system~\eqref{eq:siso} with an initially unstable plant parameters $a=1.02$, $b=1.0$, and a bounded disturbance sequence $|w_k|\le0.1$ over a finite horizon. Figure~\ref{fig:siso_cl} shows output trajectories, with an untuned gain $\kappa_0$, the output $y_k$ exhibits large excursions, while with a tuned gain $\kappa^\star$, it yields trajectories that remain within the analytical ISS bound $\pm w_{\max}$ implied by the stability contract.  Figure~\ref{fig:siso_train} shows the evolution of the gain $\kappa$ and the corresponding contract residual $C_{\mathrm{stab}}$: the gain is iteratively updated by gradient steps and then projected onto the admissible interval $\{|a-b\kappa|\le 1-\epsilon\}$, with residual remaining strictly within nonpositive values, certifying that all iterates satisfy the stability contract during training.

\subsection{Example 2: Joint Learning of a Neural Policy and Lyapunov Certificate}
\label{sec:example-node-lyap}

The second example illustrates how the $\partial$CBD framework can be used to jointly learn a neural feedback policy and a neural Lyapunov function that provides a trajectory-level stability certificate for a nonlinear closed-loop system.
In this example, we demonstrate the universality of the proposed $\partial$CBD framework by reproducing the differentiable predictive control (DPC) methodology~\cite{Drgona2024,DRGONA202280,Mukherjee2022}.

\paragraph{$\partial$CBD representation.}
We adopt a standard controlled Van der Pol oscillator as the plant dynamics
\begin{equation}
\label{eq:vdp-cont}
\dot{x}_1(t) = x_2(t), \quad
\dot{x}_2(t) = \mu\big(1 - x_1(t)^2\big)x_2(t) - x_1(t) + \Uvec(t),
\end{equation}
with state $x = [x_1,x_2]^\top \in \mathbb{R}^2$, input $u \in \mathbb{R}$, and parameter $\mu>0$.  
In the $\partial$CBD formalism, the plant is represented as a stateful block $\calB_P$ with input $U_P=\{u\}$, state $X_P=\{x\}$, output $Y_P=\{y\}$, sampling period $T_P=\{\tau\}$, and discrete-time transition and output maps
\begin{subequations}
\label{eq:vdp-disc}
\begin{align}
x_{k+1} &= f_P^{(\tau)}(x_k,u_k;\mu), \\
y_k &= g_P(x_k,u_k) = x_k,
\end{align}
\end{subequations}
where $f_P^{(\tau)}$ is the numerical integration of~\eqref{eq:vdp-cont} over one sampling period (here implemented using RK4 integrator).

The feedback policy is represented as a stateless controller block $\calB_\pi$ with inputs $U_\pi = \{x,r\}$, output $Y_\pi = \{u\}$, and algebraic map
\begin{equation}
u_k = g_\pi(x_k,r_k;\Theta)
= \pi_\Theta(x_k,r_k),
\end{equation}
where $\pi_\Theta$ is a neural network with parameters $\Theta$, taking the current state $x_k$ and reference $r_k$ as inputs and producing a bounded control signal $u_k\in[u_{\min},u_{\max}]$ via a final projection layer.

To certify stability, we introduce a Lyapunov block $\calB_V$ with input $U_V=\{x\}$ and output $Y_V=\{V\}$, implementing a scalar Lyapunov candidate $V_\phi:\mathbb{R}^2\to\mathbb{R}_+$ defining the algebraic output
$ g_V(x_k;\phi) = V_\phi(x_k)$. Our implementation follows~\cite{NEURIPS2019_0a4bbced}, where $V_\phi$ is realized as an input-convex neural network (ICNN) wrapped by a positive definite layer, ensuring $V_\phi(0)=0$ and $V_\phi(x)>0$ for $x\neq 0$.

Serial composition of $\calB_\pi$, $\calB_P$, and $\calB_V$, followed by feedback of the measured state into the policy, yields the closed-loop CBD
\begin{equation}
\calB_{\mathrm{cl}}^{\mathrm{vdp}} =\;\fbsymbol\!\big(\calB_\pi \; ;\; \calB_P \; ;\; \calB_V\big),
\end{equation}
which, for a regulation task $r_k \equiv 0$, generates trajectories
with state $X=\{x_k\}$, control 
$U=\{u_k\}$, and Lyapunov values $V=\{V_\phi(x_k)\}$.

\paragraph{Lyapunov stability contract.}
For the disturbance-free closed-loop system, asymptotic stability of the origin can be certified by a Lyapunov function $V_\phi$ satisfying a discrete-time Lyapunov inequality; here we enforce this condition as a residual certificate  
\begin{equation}
V_\phi(x_{k+1}) - V_\phi(x_k)
\;\le\; -\varepsilon,
\qquad \varepsilon>0,
\label{eq:disc-lyap-ineq}
\end{equation}
along finite horizon trajectories of~\eqref{eq:vdp-disc}.  
We encode this as a  contract attached to the block $\calB_V$, defining the Lyapunov contract residual
\begin{equation}
\calC^{\text{res}}_{\mathrm{lyap}}(X;\phi)
\;=\;
\max_{k\in\calT}
\Big( V_\phi(x_{k+1}) - V_\phi(x_k) + \varepsilon \Big),
\label{eq:vdp-lyap-contract}
\end{equation}
for a finite horizon $\calT=\{0,\dots,N\}$. The contract is satisfied if and only if
\(
\calC^{\text{res}}_{\mathrm{lyap}}(X;\phi)\le 0,
\)
which ensures that~\eqref{eq:disc-lyap-ineq} holds at every step.  
This residual contract provides a sufficient trajectory-level certificate rather than a universally quantified Lyapunov guarantee.

\paragraph{Contract-guided joint policy and certificate learning.}
Let $\theta=(\Theta,\phi)$ collect both policy and Lyapunov parameters. Executing the closed-loop block $\calB_{\mathrm{cl}}^{\mathrm{vdp}}(\theta)$ from an initial condition $X_0$ over horizon $\calT$ yields trajectories $(X,U,V) = \mathrm{Exec}(\calB_{\mathrm{cl}}^{\mathrm{vdp}}(\theta),X_0)$.
We train the policy and Lyapunov networks jointly by solving the $\partial$CBD-constrained optimization
\begin{subequations}
\label{eq:ex2-node-opt}
\begin{align}
\min_{\theta} \quad
& \mathcal{J}(\theta)
:= \sum_{k\in\calT} \|x_k\|^2 , \\
\text{s.t.}\quad
& (X,U,V) \in \mathrm{Exec}\big(\calB_{\mathrm{cl}}^{\mathrm{vdp}}(\theta), X_0\big), \\
& C_{\mathrm{lyap}}(X;\phi) \;\le\; 0, \\
& u_{\min} \le u_k \le u_{\max}, \quad \forall k\in\calT,
\end{align}
\end{subequations}
where  $(u_{\min},u_{\max})$ denote  input bounds.  
We solve the problem~\eqref{eq:ex2-node-opt} by using the penalty formulation 
for the Lyapunov contract, while using projection for the input constraints.
 The resulting loss function is minimized by gradient-based updates on $\theta$ using automatic differentiation through the entire CBD, including the numerical integrator, controller, and Lyapunov blocks.

\begin{figure}[ht]
    \centering
\includegraphics[width=0.99\columnwidth]{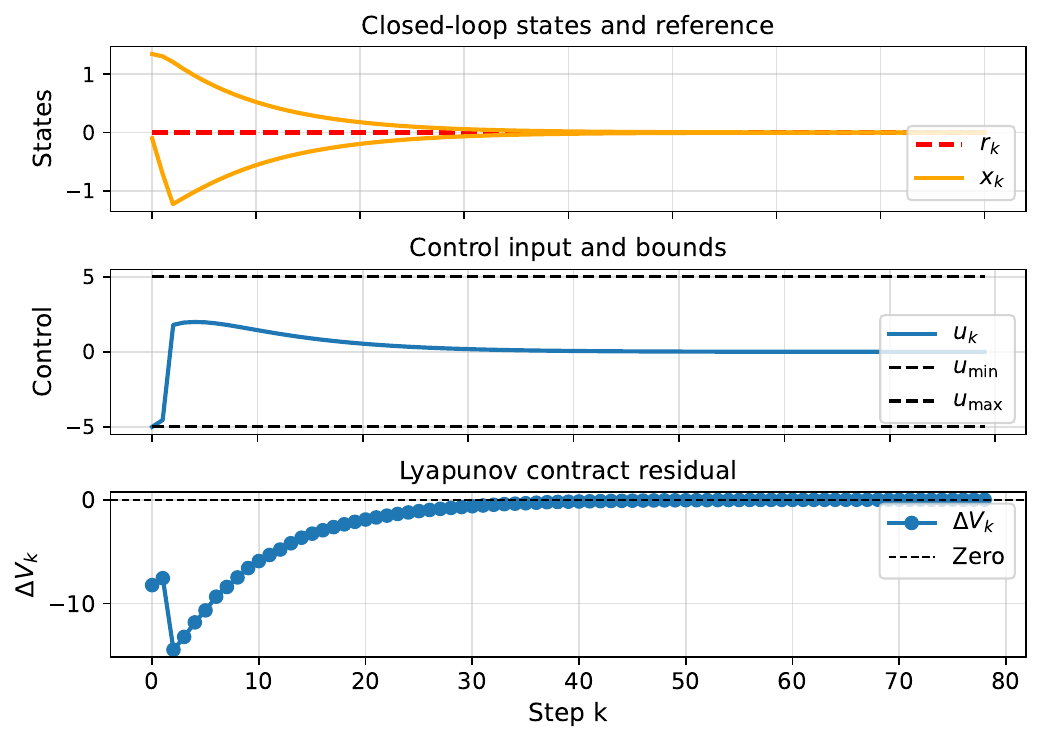}
        \vspace{-0.4cm}
    \caption{Closed-loop evolution of controlled Van Der Pol system with neural controller and Lyapunov contract.}
    \Description{Closed-loop evolution of controlled Van Der Pol system with neural controller and Lyapunov contract.}
    \label{fig:vdp_cl}
     \vspace{-0.2cm}
\end{figure}

\paragraph{Simulation results.}
Following the DPC methodology~\cite{Drgona2024,Mukherjee2022}, we train the joint policy--Lyapunov pair using batches of random initial conditions $x_0$ drawn from a compact set around the origin and a fixed horizon $N$. After training, we evaluate the learned closed-loop CBD on unseen initial conditions. Figure~\ref{fig:vdp_cl} shows the resulting closed-loop trajectories, where both states $x_k$ converge to the origin under the learned neural controller, the control input $u_k$ remains within the imposed bounds, and the Lyapunov contract residual $\Delta V_k := V_\phi(x_{k+1}) - V_\phi(x_k)$ stays strictly negative, indicating contract satisfaction along the trajectory.  
%
%
%

\subsection{Example 3: Deep Koopman Operator Learning with Stability Certificate}
\label{sec:example-deepko}

This example shows how the $\partial$CBD formalism enables contract-guided
learning of a Deep Koopman Operator (DeepKO)~\cite{Lusch2018,Folkestad2020,Yeung2019,NIPS2017_3a835d32} with explicit guarantees on
the stability of the latent dynamics.

\paragraph{$\partial$CBD representation.}
We consider a dataset of $m$ measured trajectories 
$\{\hat y_k^i\}_{k=0}^N$, $i=1,\ldots,m$, where $\hat y_k^i\in\mathbb{R}^{n_y}$ denotes the
observable at time index $k$.  
The DeepKO is represented by three blocks
\[
\calB_{\mathrm{enc}}: \hat y_k \mapsto x_k,\qquad
\calB_{K}: x_k \mapsto x_{k+1},\qquad
\calB_{\mathrm{dec}}: x_{k+1} \mapsto  y_{k+1},
\]
mapping observables $\hat y_k$ into latent coordinates $x_k\in\mathbb{R}^{n_x}$,
evolving them linearly via $K$, and decoding predictions $y_{k+1}$.

The full $\partial$CBD is the serial composition
\[
\calB_{\mathrm{koop}}(\theta)
=
(\calB_{\mathrm{enc}}(\theta_1) \,;\,
\calB_{K}(\theta_2) \,;\,
\calB_{\mathrm{dec}}(\theta_3)),
\qquad 
\theta=(\theta_1,\theta_2,\theta_3)\text.
\]
Executing the diagram from the initial data $ \hat y_0^i$ generates latent and output
trajectories
\[
(X^i,Y^i)
=
\mathrm{Exec}\!\left(\calB_{\mathrm{koop}}(\theta),\,\hat Y_0^i\right),
\qquad
k\in\calT=\{0,\ldots,N\}.
\]
The Koopman block satisfies a discrete-time stability contract
\begin{equation}
\label{eq:stability_KO}
    C_{\mathrm{stab}}(K) = 
\rho(K)<1,
\end{equation}
where $\rho(K)$ denotes the spectral radius.

\paragraph{Contract-guided identification.}
Following the general formulation \eqref{eq:verifiable-opt}, the DeepKO learning
problem is
\begin{subequations}
\label{eq:ex3-deepko-opt}
\begin{align}
\min_{\theta}\quad
& \mathcal{J}(\theta)
:= 
\sum_{i=1}^m 
\big(
\ell_y^i
+\ell_{\mathrm{lin}}^i
+\ell_{\mathrm{recon}}^i
\big),
\\
\text{s.t.}\quad
& (X^i,Y^i)
\in
\mathrm{Exec}\!\left(\calB_{\mathrm{koop}}(\theta),\,\hat Y_0^i\right),
\qquad i=1,\ldots,m,
\\
& C_{\mathrm{stab}}\!\left(K(\theta_2)\right) \le 0 .
\end{align}
\end{subequations}

The identification losses are
\begin{subequations}
\begin{align}
\ell_y^i
&=
\sum_{k=0}^{N-1}
Q_y\,\|y_{k+1}^i - \hat y_{k+1}^i\|_2^2, \\
\ell_{\mathrm{lin}}^i
&=
\sum_{k=0}^{N-1}
Q_x\,
\big\|
\calB_{\mathrm{enc}}(\hat{y}_{k+1}^i)
- 
\calB_{K}^k(\calB_{\mathrm{enc}}(\hat{y}_k^i))
\big\|_2^2, \\
\ell_{\mathrm{recon}}^i
&=
Q_{\mathrm{recon}}
\big\|
y_0^i
-
\calB_{\mathrm{dec}}(\calB_{\mathrm{enc}}(y_0^i))
\big\|_2^2.
\end{align}
\end{subequations}

\paragraph{Stability contract.}
To ensure $\rho(K)<1$, we parameterize $K=U\Sigma V^\top$ via Singular Value Decomposition (SVD), constrain the
singular values in $\Sigma$ through a sigmoid-based scaling, and enforce the
orthogonality of $U,V$ using the Householder reflectors~\cite{Zhang2018StabilizingGF}.
This satisfies the stability contract by construction.
In contrast to Example~2, this constraint is enforced at the parameter level and
constitutes an exact A--G stability contract for the latent linear
dynamics, rather than a trajectory-level certificate.

\begin{figure}[h]
    \centering
    \begin{subfigure}{0.49\columnwidth}
        \centering
        \includegraphics[width=\linewidth]{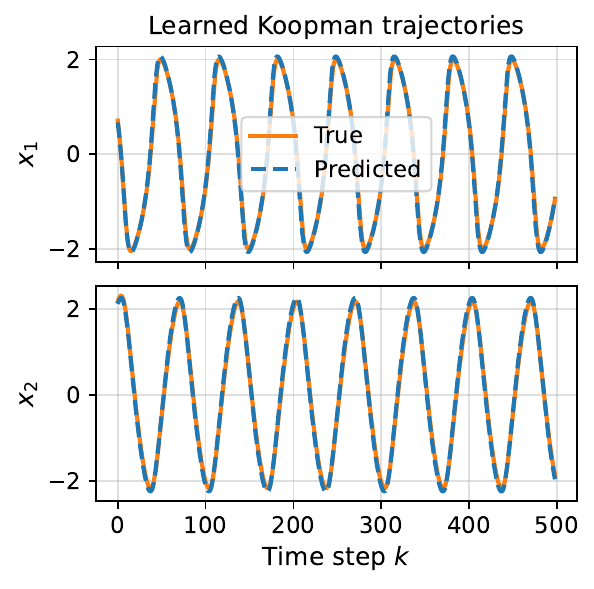}
        \caption{True vs.\ predicted  trajectories.}
        \Description{Time series plot comparing true Van der Pol trajectories with predictions of the learned Deep Koopman model over 500 steps.}
        \label{fig:koopman_traj}
    \end{subfigure}
    \hfill
    \begin{subfigure}{0.49\columnwidth}
        \centering
        \includegraphics[width=\linewidth]{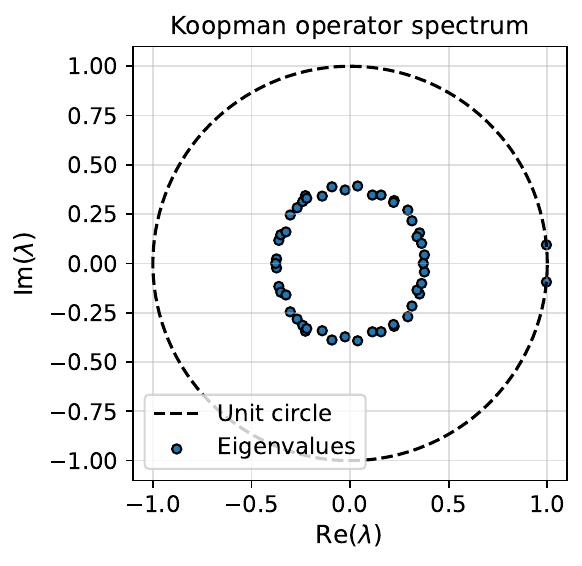}
        \caption{Koopman spectrum.}
        \Description{Scatter plot of Koopman eigenvalues in the complex plane overlaid with the unit circle, showing all eigenvalues inside the unit disk.}
        \label{fig:koopman_eigs}
    \end{subfigure}
    \vspace{-0.2cm}
    \caption{Learned Deep Koopman model for the Van der Pol oscillator: time–domain rollouts (left) and spectrum of the stabilized Koopman operator (right).}
    \Description{Two side-by-side plots: left shows true and predicted Van der Pol trajectories; right shows eigenvalues of the learned Koopman operator within the unit circle.}
    \label{fig:koopman}
\end{figure}

\paragraph{Simulation results.}
We train the Deep Koopman CBD with stability contract~\eqref{eq:stability_KO} by solving the problem~\eqref{eq:ex3-deepko-opt} with training trajectories of the uncontrolled Van der Pol oscillator~\eqref{eq:vdp-cont}. After training, we evaluate the model on an unseen initial condition over a $ 500$-step rollout. Figure~\ref{fig:koopman} (left) compares the true state trajectories with the Koopman predictions, showing close agreement over the entire horizon. Figure~\ref{fig:koopman} (right) displays the eigenvalues of the learned Koopman operator in the complex plane together with the unit circle; all eigenvalues remain inside the unit disk, certifying that the learned linear representation is contractive and consistent with the imposed stability contract.

\section{Conclusions}

This paper introduced $\partial$CBDs, a differentiable extension of causal block diagrams that unifies three traditionally disjoint paradigms in cyber-physical systems engineering: (i) compositional modeling via CBDs, (ii) formal correctness via assume--guarantee and residual contracts, and (iii) learning and optimization via differentiable programming. By equipping blocks with explicit interfaces, time semantics, and contract annotations that admit both universal A--G semantics and differentiable residual certificates, $\partial$CBDs transform CBDs from a simulation-oriented formalism into a trainable, verifiable, and compositional modeling framework. The resulting diagrams are simultaneously causal, certifiable, and gradient-compatible, enabling end-to-end learning of feedback systems with embedded safety and performance guarantees.

Looking ahead, $\partial$CBDs provide a foundation for verifiable agentic AI, enabling perception--planning--control loops with differentiable contract layers that support data-driven adaptation while preserving correctness. Their compositional contracts naturally support scalable certification of large, multi-rate, and multi-agent systems. The framework also enables hybrid intelligence architectures that combine symbolic reasoning, optimization layers, and neural components under explicit contracts for interpretable and safe operation. More broadly, $\partial$CBDs bridge modern automatic differentiation libraries with safety-critical verification toolchains, supporting real-time learning with formal runtime assurance.


\begin{acks}
This work was partially supported by the National Science Foundation under Award No.~2514584.
J\'an Drgo\v{n}a was supported by the Ralph O’Connor Sustainable Energy Institute at Johns Hopkins University.
\end{acks}

\printbibliography
\end{document}